# Broadband terahertz comb with sub-Hz comb linewidth

Min Li,[a,†,*] Yu Xia,[b,†] Yuan Chen,[b,†] Sinan Tao,[a] Mingyang He,[c] Kun Huang,[b] Hua Li,[d] Ming Yan,[b,*] and Heping Zeng[b,c,e,*]

[a]School of Optical Electrical and Computer Engineering, University of Shanghai for Science and Technology, Shanghai, China

[b]State Key Laboratory of Precision Spectroscopy, and Hainan Institute, East China Normal University, Shanghai, China

[c]Jinan Institute of Quantum Technology, Jinan, China

[d]Key Laboratory of Terahertz Solid State Technology Shanghai Institute of Microsystem and Information Technology, Chinese Academy of Sciences, Shanghai, China

[e]Chongqing Key Laboratory of Precision Optics, Chongqing Institute of East China Normal University, Chongqing, China

[†]These authors contributed equally to this work.

[*]Address all correspondence to Min Li, minli1220@usst.edu.cn; Ming Yan, myan@lps.ecnu.edu.cn; Heping Zeng, hpzeng@phy.ecnu.edu.cn

**Abstract:**

Terahertz (THz) frequency combs are increasingly essential for spectroscopy, metrology, and quantum science. However, generating a dense array of evenly spaced ultra-narrow THz comb lines is challenging. Here, we demonstrate broadband THz comb generation using a photoconductive antenna that transfers a noise-suppressed near-infrared electro-optical (EO) comb into the THz domain. Our noise-suppression strategy, leveraging soliton self-frequency shift and spectral filtering, effectively suppresses EO comb phase noise without requiring active stabilization. The resulting THz comb exhibits broad spectral coverage (0.05-4 THz), narrow comb linewidths (0.3 Hz at the Fourier-transform limit), and excellent frequency stability ($8.6 \times 10^{-14}$ at 1-second integration). We further demonstrate asynchronous THz time-domain spectroscopy, resolving ~36,000 comb lines with 50 MHz spacing. Crucially, the inherent frequency agility of the EO comb enables rapid and wide-range tuning of the THz comb line spacing. These attributes position our THz comb as a versatile tool for high-resolution molecular spectroscopy and precision THz metrology.

# Introduction

Frequency comb synthesis in the terahertz (THz) domain (~0.1 to 10 THz) is crucial for advancing THz technologies across various applications, from molecular spectroscopy[1] to THz metrology.[2] Advanced THz combs promise to generate numerous evenly spaced, ultra-stable frequency lines traceable to atomic clocks.[3] This capability is highly desirable for high-resolution molecular spectroscopy,[4] enabling precise identification of complex molecules such as biomolecules and atmospheric trace gases, with applications including environmental monitoring,[5] medical diagnostics (e.g., THz breath analysis for disease detection),[6] and fundamental studies of molecular structures.[4] Beyond spectroscopy, ultra-stable THz combs find applications in quantum science, where they can be used to probe Rydberg states,[7] superconducting qubits,[8] and cold molecular gases,[9] advancing quantum computing and precision timekeeping. In precision metrology, these THz combs enable direct THz-to-optical clock comparisons,[3] linking THz frequencies to SI-traceable optical standards. Despite their importance, developing metrological-grade THz combs with both broad bandwidth and narrow comb linewidths remains challenging.[10]

To date, most metrological THz sources operate in a single mode, typically within a limited tuning range, e.g., 0.1-1.1 THz,[11] despite offering sub-Hz linewidth and fractional frequency stability between $10^{-12}$ and $10^{-15}$ at 1 s, such as those achieved by phase-locking QCLs to a free-space THz comb[12,13] or by photomixing continuous-wave (CW) lasers stabilized to a near-infrared laser comb.[14,15] Notably, fully stabilized, multi-mode THz QCL combs have achieved relative line frequency accuracy near $10^{-12}$.[16] In general, THz QCLs provide only a limited number of lines (typically a few tens) with relatively large line spacing (e.g., 17.45 GHz),[17] but deliver high per-comb-line power, which is advantageous for applications such as probing water molecules that exhibit strong absorption in the THz regime.[18] By contrast, many high-resolution spectroscopic applications,[4] as discussed above, demand densely spaced broadband THz combs.

Broadband THz combs with dense comb lines can be generated through optical-to-THz conversion using photoconductive antennas (PCAs)[19,20] or electro-optic crystals,[21,22] with the THz bandwidth dictated by the optical pulse duration. Mode-locked femtosecond (fs) lasers produce THz radiation spanning, for example, 0.1-4 THz,[23,24] enabling broadband, comb-line-resolved THz spectroscopy, aided by repetition-frequency stabilization[25] or adaptive phase correction.[26-28] However, the resulting THz frequency accuracy has been limited to $10^{-9}$, corresponding to kHz comb linewidths.[25,28] Further improvements require tightly stabilizing an fs laser comb to an optical reference, such as an ultra-stable CW laser.[29] However, this approach faces challenges in suppressing the fs comb's wideband phase noise (due to the limited electronic bandwidth of phase-lock loops)[30] and significantly increases system complexity, reducing practicality. Additionally, mode-locked cavities impose limitations on THz comb line-spacing tunability (~1% of the comb repetition frequency),[30] restricting their use in demanding applications, such as resolving hyperfine transitions of gas molecules.[31]

Electro-optical (EO) combs offer an alternative approach to THz wave generation, providing high passive frequency stability inherited from an atomic clock, wide tunability in both line frequency and spacing, and a simple, robust architecture.[32,33] Recent advancements in EO comb technology include octave-spanning EO comb generation,[34] on-chip EO synthesis,[35] and extensions into the ultraviolet[36] and mid-infrared.[37] However, their application in the THz regime remains constrained by narrow bandwidth (~100 GHz)[38] and low comb line density (typically a few dozen).[39] Expanding THz bandwidth requires a broadband fs EO comb, typically achieved via nonlinear spectral broadening,[33] but this introduces excess noise, widening comb linewidths and degrading coherence.[32]

In this paper, we demonstrate metrological-grade, broadband THz comb synthesis using a PCA driven by a low-noise, near-infrared fs EO comb. Innovatively, we introduce a passive phase noise suppression strategy based on soliton self-frequency shift (SSFS) and spectral filtering. This strategy reduces wideband phase noise of the EO comb without the use of phase-locking electronics. Consequently, we generate a THz comb with low phase noise (-80 dBc/Hz), exceptional frequency stability ($8.6 \times 10^{-14}$ at 1-s integration), and sub-Hz comb linewidths (limited only by measurement time). Using asynchronous THz time-domain spectroscopy (TDS), we demonstrate a broad THz spectrum spanning 0.05-4 THz with ~36,000 resolved comb lines (within a 45 dB dynamic range). Our frequency-agile THz comb, operating without mode-locked cavities, enables rapid and arbitrary tunability of comb line spacing, making it a powerful tool for various spectroscopic and metrological applications.

# Results

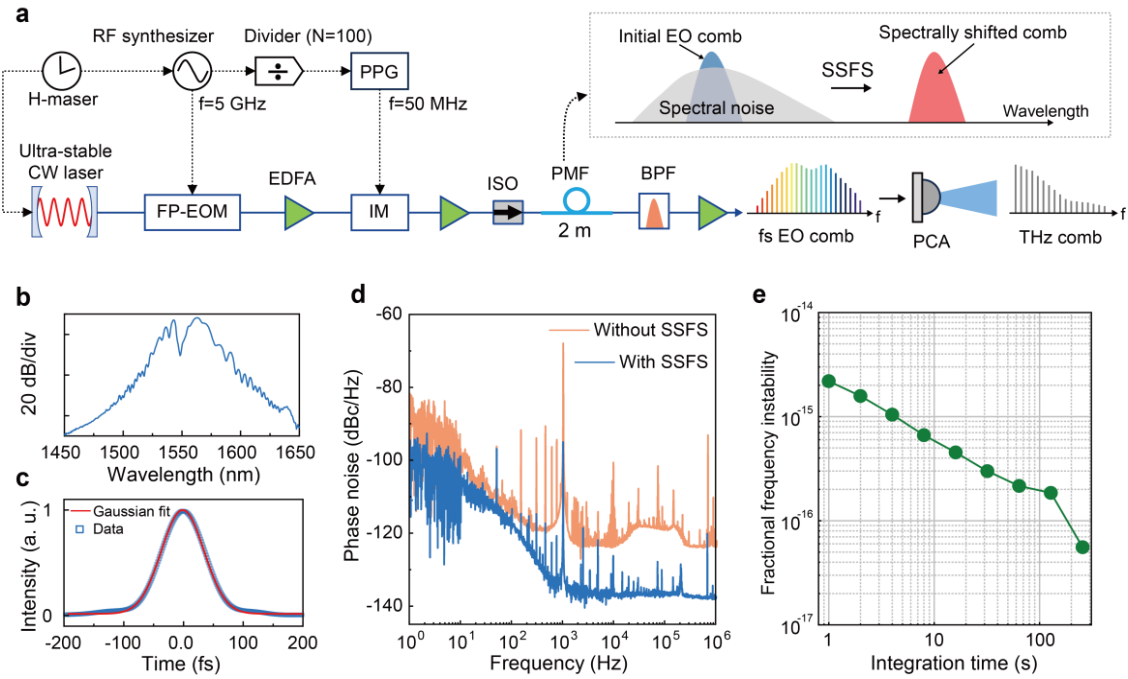

**Fig. 1 Terahertz (THz) comb generation. a** Experimental setup. CW: continuous-wave; RF: radio-frequency; H-maser: Hydrogen maser; PPG: picosecond pulse generator; FP-EOM: Fabry-Pérot electro-optical modulator; EDFA: erbium-doped fiber amplifier; IM: intensity modulator; ISO: fiber-coupled isolator; PMF: polarization-maintaining fiber; BPF: band-pass filter; PCA: photoconductive antenna; SSFS: soliton self-frequency shift; f: frequency; N: number of divisor. **b** and **c** Spectrum and autocorrelation trace of the femtosecond (fs) EO comb. The autocorrelation data (blue) are fitted by a Gaussian curve (red). **d** Phase noise spectra of the EO comb, measured with and without SSFS and spectral filtering. **e** Allan deviation of the EO comb. The cartoon in the dashed box illustrates the spectral separation (using SSFS) between the initial EO comb and the amplified spontaneous noise, achieved by SSFS in a 2-m PMF.

Our experimental scheme is illustrated in Fig. 1a. A compact, frequency-stabilized CW laser (linewidth < 1 Hz at 1550 nm, SLS-INT-1550-200, SLS) feeds a Fabry-Pérot EO modulator (FP-EOM), generating a near-infrared comb with 5 GHz line spacing. The line spacing is set by a radio-frequency (RF) synthesizer disciplined to a hydrogen maser with a frequency stability of $10^{-13}$ in 1s. After the FP-EOM, an intensity modulator (IM) is employed as a pulse picker, driven by a picosecond pulse generator (PPG). The PPG operates at one-hundredth of the fundamental frequency, enabling the IM to select one out of every 100 pulses, thereby reducing the comb repetition rate (and hence the line spacing) to $f_r$=50 MHz. Further details are provided in Methods. Notably, the reduction of our comb's line spacing to below 100 MHz serves two purposes: to match the PCA excitation bandwidth (<100 MHz) and to provide sufficient single-pulse energy for SSFS and nonlinear spectral broadening.

Due to the low duty cycle (1%), the pulse picker output power is below 0.1 mW. It is then amplified to 150 mW using a home-built erbium-doped fiber amplifier (EDFA). During this process, amplified spontaneous emission (ASE) arises, spectrally overlapping with the EO comb. Their mixing on a PCA will generate a significant amount of noise, compromising the stability of the THz comb. To mitigate this, we spectrally shift the comb pulse to 1575 nm—beyond the ASE spectral range (1520-1570 nm)—by harnessing SSFS in a 2-m polarization-maintaining fiber. The shifted comb light is then filtered using a bandpass optical filter. Before being launched to the PCA, the comb undergoes further power amplification and spectral broadening (Fig. 1b) in an EDFA, followed by temporal compression to 65 fs (Fig. 1c) via dispersion management in fibers. Autocorrelation traces measured at different stages of the setup are shown in Supplementary Fig. S1.

To suppress ASE noise, we employ the SSFS-based noise suppression scheme (shown in a dashed box in Fig. 1a). In this scheme, EO comb pulses shift in wavelength within a 2-m fiber due to nonlinear effects like stimulated Raman scattering. Because the shift depends nonlinearly on pulse energy, the comb pulses—having much higher energy than ASE noise—are spectrally separated from the ASE background. Bandpass filtering then yields low-noise comb pulses. Note that standard bandpass filtering cannot isolate ASE that lies within the filter passband. Our approach circumvents this limitation by spectrally shifting the near-infrared comb into a low-noise spectral window, thereby effectively eliminating parasitic ASE. Importantly, this scheme offers two advantages over conventional comb systems: (1) it eliminates the need for complex phase stabilization systems;[11] and (2) it effectively suppresses wideband noise, which poses a challenge for existing active stabilization systems due to their limited operation bandwidth (typically 10-100 kHz).[40] A numerical simulation regarding this SSFS-based noise suppression scheme is provided in Supplementary Material Fig. S2 and Supplementary Material Note 1.

Experimentally, we measure phase noise spectra before and after SSFS (see Methods). The results, plotted in Fig. 1d, demonstrate wideband noise reduction in an RF range from 10 Hz to >1 MHz (i.e., approximately 10 dB in the frequency range below 10 Hz and ~20 dB for frequencies above 1 kHz). Notably, this leads to a low phase noise level, reaching -136 dBc/Hz at 1 kHz. Furthermore, we measure the EO comb's fractional frequency instability to be $10^{-15}$ at 1-s integration time, which improves to $5.57 \times 10^{-17}$ at 256 s integration (Fig. 1e). This is a remarkable achievement, considering the setup does not involve complicated active stabilization systems,[11] apart from the cavity-stabilized CW seed laser. Although systems utilizing a sub-Hz CW laser and FP-EOM have been demonstrated for octave-spanning EO comb generation,[33,34] our setup is distinguished by its low phase noise passively achieved by SSFS. More importantly, we demonstrate, for the first time, its application to broadband, ultra-stable THz comb generation.

We generate THz comb radiation by fiber-coupling the EO comb to the PCA. To characterize the THz comb line's frequency stability, we perform heterodyne detection between our THz comb and an ultra-stable CW THz source. The setup is outlined in Fig 2a. The beat signal is digitized in the time domain at a sampling rate of 10 MS/s and subsequently Fourier-transformed into an RF spectrum. As an example, a recorded time-domain trace is shown in Fig. 2b, with its corresponding RF spectrum in Fig. 2c. The total recording time is 3.146 s, limited by the digitizer's memory depth. The -3dB linewidth of the beat note is 0.3 Hz (inset of Fig. 2c), limited by the RF spectral resolution.

We then characterize the beating signal's phase noise spectrum using a commercial spectrum analyzer. The recorded noise power density spectrum is presented in Fig. 2d, exhibiting an exceptional low noise level of -80 dBc/Hz at 1 kHz offset. Furthermore, to evaluate the fractional frequency instability of the beating signal, we monitor it over a 30-minute period with a digital frequency counter and perform an Allan deviation analysis. The results are plotted in Fig. 2e. The fractional frequency instability ($=\sigma_{SD}/f_{cw}$) is $8.6 \times 10^{-}$ [14] at a 1-s integration time, where the standard deviation ($\sigma_{SD}$) of the frequency jitter is 8.3 mHz. The frequency stability improves to $5.1 \times 10^{-15}$ at 265 s, which is three orders of magnitude better than that of existing PCA-based broadband THz comb systems.[25,28] Details of these measurements are provided in Methods.

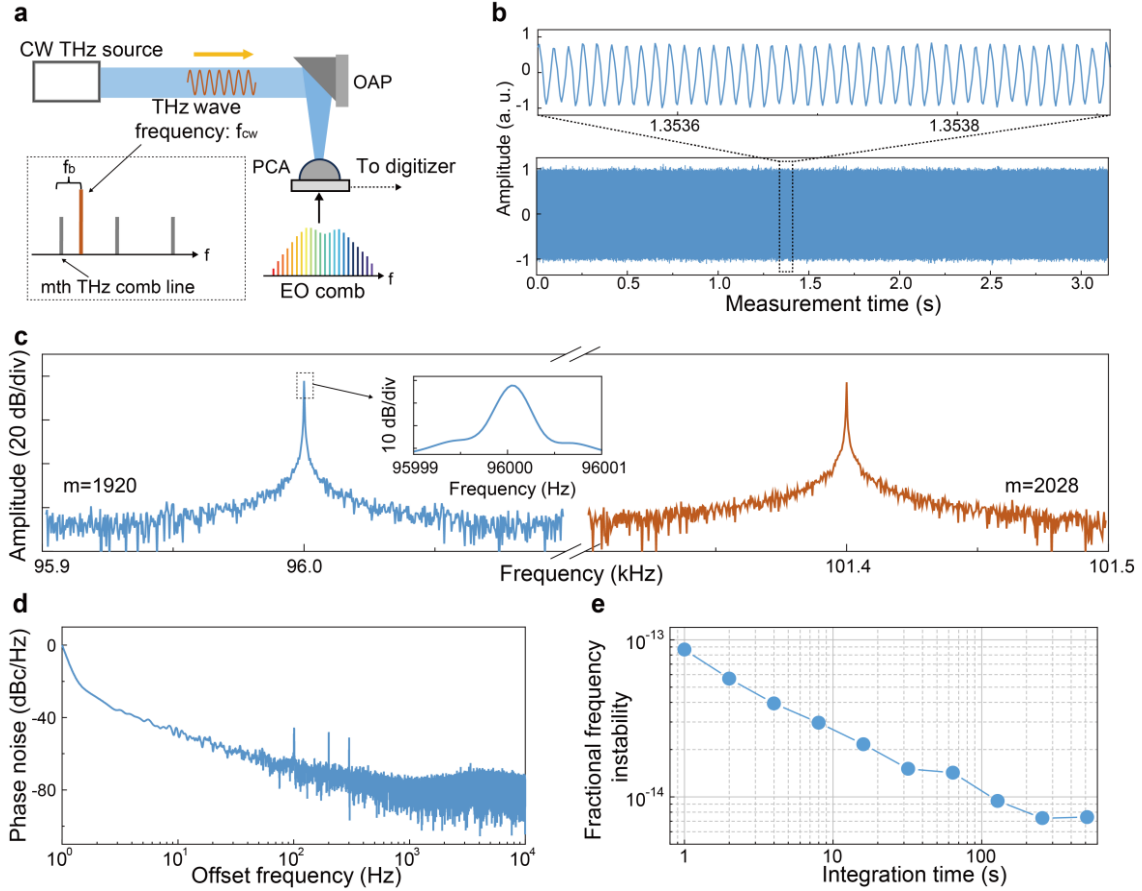

**Fig. 2 Results of comb linewidth and frequency stability. a** Schematic of heterodyne detection between the THz comb and an ultra-stable continuous-wave (CW) source. EO: electro-optical; PCA: photoconductive antenna; OAP: off-axis parabolic mirror. Here, $f_{cw}$ is the frequency of the CW radiation, $f_b$ is the beat frequency, and f represents the general frequency. **b** Beating signals in the time domain. **c** Fourier-transformed beat-note spectra, where m denotes the comb line index. **d** Phase noise spectrum of a beat note. **e** Allan deviation of a beat frequency.

In the heterodyne measurement, the beating frequency ($f_b$) between the CW THz radiation (frequency: $f_{cw}$=96 GHz) and the $m$th comb line ($f_{comb, m}$) can be expressed as $f_b$ =|$f_{comb, m}$ -$f_{cw}$|, where $f_{comb, m} = mf_r$ and $m$ is an integer. Since our THz comb's offset

frequency is zero,[26] we deduce that $m$=1920. By tuning $f_{cw}$ to 101.4 GHz, we measure the beating signal (Fig. 2c) for the comb line at $m$=2028, which exhibits the same Fourier-transform-limited linewidth (0.3 Hz). Our THz comb, with excellent frequency stability and sub-Hz comb linewidths, provides an absolute frequency calibration tool for metrological applications. For example, based on the Vernier principle,[3] our THz comb can precisely measure the absolute frequency of any THz wave within the comb frequency range. Examples are provided in Supplementary Material Fig. S3.

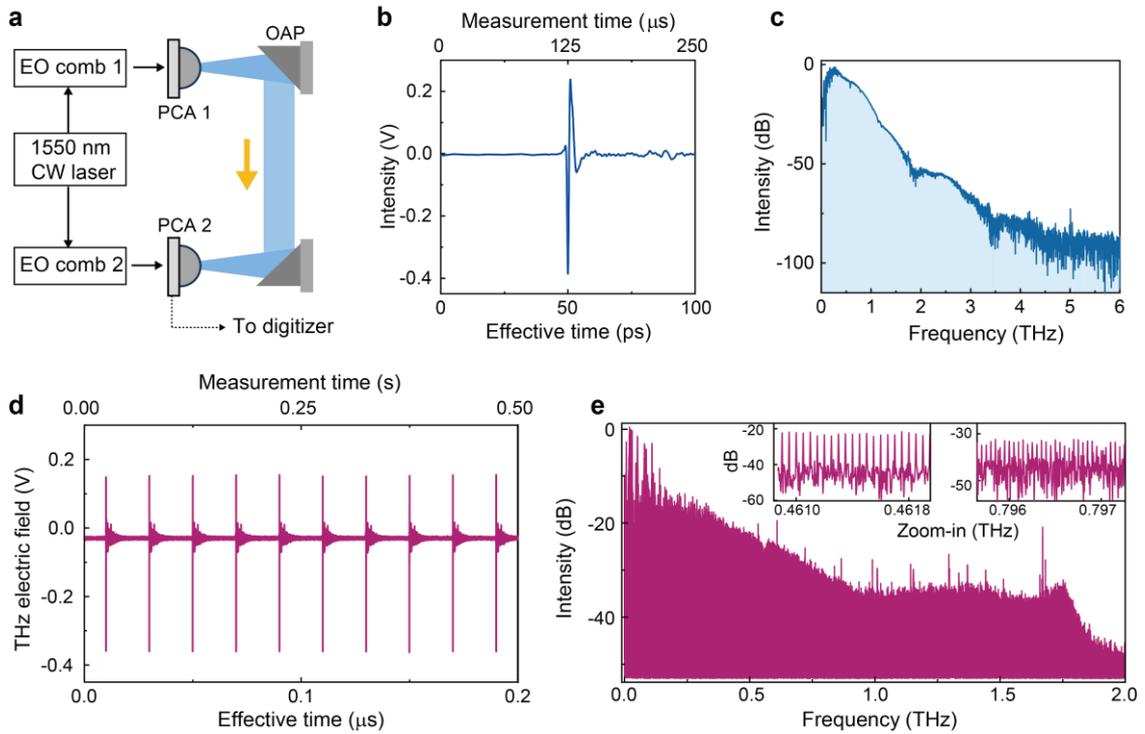

**Fig. 3 Time-domain spectroscopy. a** Experimental schematic. EO comb 1 and EO comb 2 are co-seeded by the same CW laser at 1550 nm. The THz radiation, generated by converting EO comb 1 into the THz domain via a photoconductive antenna (PCA), is guided to a second PCA through a pair of off-axis parabolic mirrors (OAPs). In the second PCA, the THz radiation is asynchronously sampled by EO comb 2. **b** Single-shot time-domain THz signal and **c** the corresponding Fourier transformed spectrum, averaged over 10 measurements. **d** A train of time-domain THz signals. **e** Comb-line-resolved THz spectrum. The insets show magnified views of the resolved comb lines.

Furthermore, we demonstrate the broadband measuring capability of our THz comb by performing asynchronous THz TDS—an analog to dual-comb spectroscopy—with widespread applications in THz spectroscopy, sensing, and imaging.[41] The experimental scheme is drawn in Fig 3a, with further details provided in Methods. In short, an EO comb (EO comb 1) hits a PCA, generating THz comb radiation, which is then guided to a second PCA acting as a receiver. Meanwhile, another fs EO comb (EO comb 2) with a slightly different repetition frequency ($f_r + \Delta f_r$) is sent to the receiver for THz field-resolved optical sampling.[30] Figure 3b shows a recorded THz field signal in both measurement time ($t_m$) and effective time ($t_{eff} = t_m \cdot \Delta f_r / f_r$). The corresponding Fourier-transformed THz spectrum, spanning 0.05-4 THz with a dynamic range of 85 dB, is plotted in Fig. 3c. This spectrum is obtained in a single measurement of 1 ms, corresponding to a refresh rate of 20 Hz (i.e., $\Delta f_r$).

An advantage of using EO comb synthesis for asynchronous THz TDS is the excellent mutual temporal coherence between the THz comb and the sampling EO comb, achieved without the need of active stabilization or post phase correction. This enables time-domain coherent averaging. For instance, Fig. 3d shows a train of stable THz signals with 200-fold averaging performed directly in the time domain, simplifying data processing. The total averaging time is 100 s. The corresponding high-resolution spectrum, shown in Fig. 3e, reveals ~36,000 discrete THz comb lines spanning 0-1.8 THz, with a uniform frequency spacing of $f_r$ (=50 MHz). The insets provide magnified views of the resolved comb lines. These lines exhibit Fourier-transform limited linewidths, i.e., 2 Hz in the RF domain and 5 MHz in the THz domain (linked by $f_r/\Delta f_r$ =2.5×10$^6$). Currently, the number of resolvable comb lines is limited by the THz comb power (~0.1 mW), which can be improved by increasing the excitation EO comb power but ultimately constrained by the PCA's damage threshold. This is a common issue in PCA-based schemes, highlighting the need for further development of high-efficiency optical-to-THz converters.[42] Nevertheless, the current THz comb and TDS system enable broadband, comb-line-resolved THz spectroscopy, with applications in multi-gas analysis and precision molecular spectroscopy.[4-6]

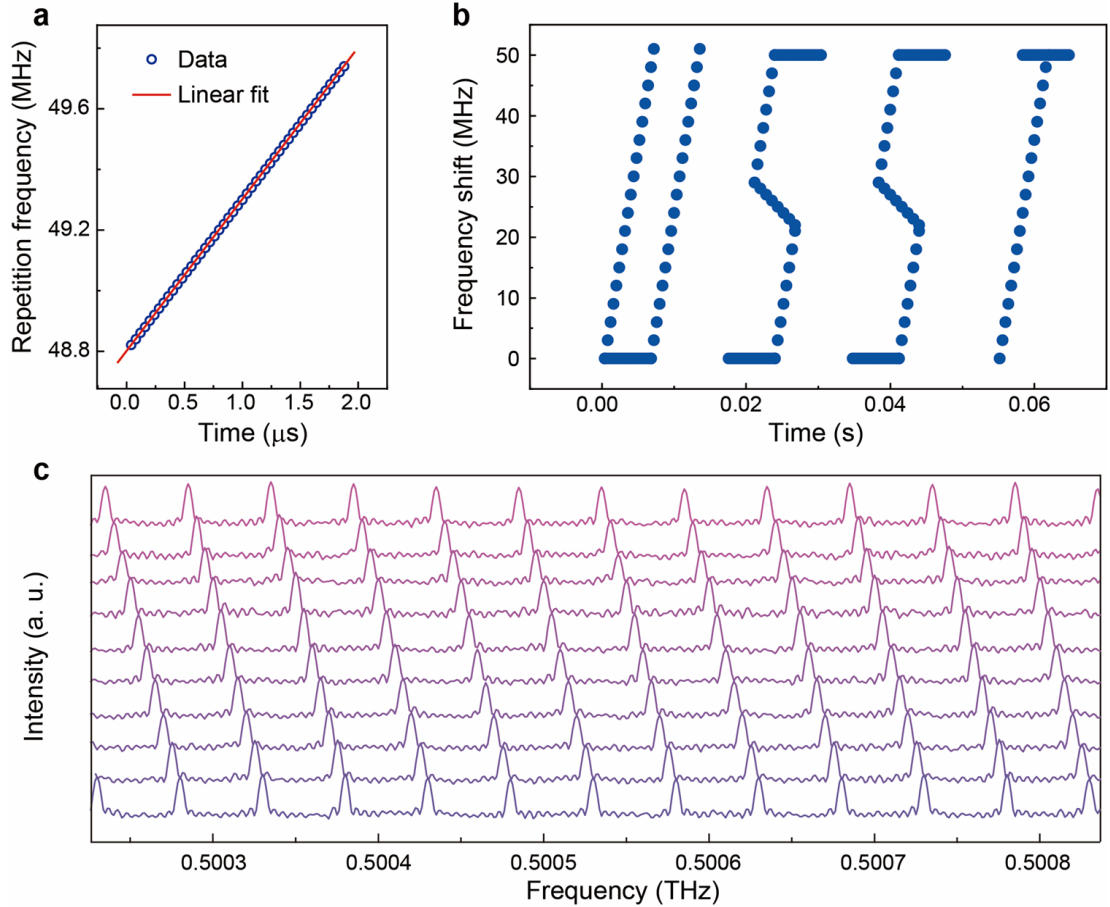

**Fig. 4 Results of tuning comb line frequencies. a** Fast tuning of an EO comb's repetition frequency. **b** Beat-note frequency shifts recorded when tuning a THz comb's repetition frequency. The abbreviation "USST" stands for the University of Shanghai for Science and Technology. c Interleaved spectra with resolved THz comb lines. In **a**, the data points are linearly fitted (red line) with $R^2 > 0.999$. In **c**, each spectrum is obtained by tuning the THz comb's repetition frequency with a step size of 500 Hz (starting at 50 MHz), resulting in a comb line frequency shift of ~5 MHz around 0.5 THz. During the tuning process, the repetition frequency of the sampling comb (EO comb 2 in Fig. 3a) is synchronously adjusted while maintaining a constant offset frequency of 20 Hz.

The capability to rapidly adjust comb line frequencies over a wide range is crucial for spectral interleaving,[43] with applications spanning Doppler-free molecular spectroscopy[44] to high-resolution gas sensing.[32] However, THz combs generated by mode-locked lasers exhibit restricted tunability in terms of tuning speed, linearity, and range, particularly when the offset frequency is null and the repetition frequency is constrained by optical cavities.[44] In contrast, our EO comb, unbound by the constraints of a mode-locking cavity, allows swift and extensive tuning of comb lines. For instance,

Fig. 4a illustrates the rapid tuning of the EO comb's repetition frequency ($f_r$ or line spacing), achieved by configuring the RF synthesizer to operate in a frequency-sweeping mode (Fig. 1a). The repetition frequency is concurrently recorded in the time domain using a high-speed oscilloscope (Agilent DSO81304B) and subsequently transformed into the RF domain through post-processing. In Fig. 4a, the $f_r$ is tuned from 48.82 to 49.74 MHz within only 2 μs, which is typically unachievable with conventional mode-locked lasers.[25-30] In this tuning range, the spectral and temporal properties of the EO comb change negligibly. Importantly, this tuning range (0.92 MHz) sufficiently bridges the 50 MHz line-spacing gap for the comb lines at $m > 54$ or $f_{comb, m} = mf_r > 2.7$ GHz. The plots in Fig. 4b show the beating frequencies between the CW THz source and one of the comb lines ($m$=1920) during tuning, demonstrating the ability to arbitrarily adjust the comb line's frequency within the 50 MHz gap.

Ultimately, the tunability of the THz comb enhances THz spectroscopy by improving spectral sampling resolution. Figure 4c presents the overlaid spectra of the resolved comb lines, spaced by ~5 MHz in the THz domain, representing a tenfold improvement in the initial spectral sampling resolution (50 MHz). In this experiment, we simultaneously tune the repetition frequencies of both the THz comb and the sampling EO comb while maintaining a constant 20 Hz offset between them. These results highlight the potential of our THz comb for high-resolution THz spectroscopy.

## Discussion and conclusion

In this work, we demonstrate a THz comb achieving simultaneous broad spectral coverage (0.05-4 THz) and narrow comb lines with a linewidth of < 0.3 Hz, three orders of magnitude finer than existing broadband THz comb sources (see Table 1). Our THz comb exhibits high accuracy in comb-line frequencies, with a fractional frequency instability of $8.6 \times 10^{-14}$ in 1 s ($5.1 \times 10^{-15}$ at 265 s), approaching the performance of state-of-the-art single-mode THz wave sources (e.g., typically ranging between $10^{-14}$ and $10^{-15}$ at 1 s).[11] Further comparisons are detailed in Supplementary Material Fig. S4.

Currently, our system faces two main limitations. The first is the limited THz comb power (approximately 0.1 mW), constrained by the damage threshold of the PCA. The second is the inherent trade-off between the number of comb lines and the available power per line. While our source generates a large number of comb lines (~36,000), the average per-line power is relatively low (~3 nW), in contrast to THz QCLs.[45] These limitations may be mitigated by advances in THz antenna architectures and the development of high-damage-threshold crystals with enhanced conversion efficiency. Nonetheless, as metrological-grade broadband THz combs are still under development, our THz source finds a range of applications. For instance, it can serve as a THz reference for stabilizing THz QCLs and generating high-power, ultra-stable THz waves for applications such as high-precision THz radar,[46] next-generation (5G/6G) wireless communications,[47] and industrial non-destructive testing.[48] As a THz frequency ruler, its broad coverage and high frequency stability enable precise measurement of unknown THz frequencies and simultaneous interrogation of multiple non-adjacent molecular lines.[49] Even in its current implementation, with an output power of 0.1 mW, our THz comb is already suitable as a calibration standard for gas-phase THz spectroscopy, including precision measurements of water vapor transitions and Rydberg atomic states. Particularly, its highly tunable line spacing further facilitates the precise measurement of dense molecular transitions,[50] essential for molecular structure determination, particle physics studies, and fundamental constant testing. Moreover, our THz comb enhances THz spectroscopic sensing by improving spectral resolution and frequency accuracy, benefiting applications such as trace gas detection,[51] environmental monitoring,[5] and human breath diagnostics.[6]

Finally, our work advances THz TDS—a powerful tool for THz sensing, screening, and imaging—by replacing bulky mode-locked lasers with a compact EO comb and eliminating the need of active stabilization systems. Also, with recent progress in on-chip EO modulators[35] and advanced THz antennas, our source can be further miniaturized for enhanced portability.

**Table 1** Comparison of THz comb sources.

| Ref. | Platforms | Spectral coverage | Number of comb lines | Comb linewidth | Comb line frequency instability |
|---|---|---|---|---|---|
| 13 | QCL combs | Full width: 66 GHz (center at 4.2 THz) | 11 | 14.8 kHz | $3.5 \times 10^{-9}$ |
| 17 | QCL combs | Full width: 275.5 GHz (center at 2.9 THz) | 19 | 2 Hz | $2 \times 10^{-12}$ |
| 25 | Mode-locked lasers + PCA | 0.15-2.4 THz | 28,125 | 3.7 kHz | $1.8 \times 10^{-9}$ |
| 28 | Mode-locked lasers + PCA | 0.05-1.4 THz | 20,411 | 1.9 kHz | $1.7 \times 10^{-9}$ |
| 39 | EO combs + PCA | Full width: 10 GHz (center at 0.096 THz) | 100 | / | $5.2 \times 10^{-11}$ |
| This work | SSFS-based EO combs + PCA | 0.01-4.0 THz | 36,000 | <0.3 Hz | $8.6 \times 10^{-14}$ |

## Methods

**Near-infrared electro-optical (EO) comb and THz comb generation.** In Fig. 1a, a narrow-linewidth CW laser at 1550 nm is fed to a Fabry-Pérot EO modulator (FP-EOM, OptoComb), generating a comb with a repetition frequency of 5 GHz. The repetition frequency is then reduced to 50 MHz using an intensity modulator (IM; KY-AMU-15-10G, Keyang Photonics; bandwidth: > 10 GHz). The IM is driven by a picosecond pulse generator (PPG; Keyang Photonics), which converts radio-frequency (RF) sine waves into 80 ps electrical pulses. Using home-made erbium-doped fiber amplifiers (EDFAs) and carefully managing the net dispersion of the fibers, we achieve 150 mW, 1.2 ps EO comb pulses before they enter a 2-m polarization-maintaining fiber (PMF) for soliton self-frequency shift (SSFS). Through SSFS, the comb pulses are spectrally shifted from 1550 nm to 1575 nm, after which they are spectrally filtered. The shifted pulses (~5 mW) are further amplified up to 100 mW in an EDFA and compressed to 65 fs using a piece of single-mode fiber. To generate a broadband THz comb, the EO comb (incident power: 25 mW) is sent to a fiber-coupled, low-temperature InGaAs/InAlAs photoconductive antenna (PCA; TERA 15-TX-FC, Menlo Systems; bias voltage: 95 V).

**THz comb-CW beat note measurement.** In this experiment (Fig. 2a), an ultra-stable CW THz radiation source is used. This source is generated by an active frequency multiplication chain (×6) driven by a microwave frequency synthesizer (WR 10, Virginia Diodes, Inc), disciplined to a hydrogen maser with a frequency stability of $10^{-13}$ at 1 s. The CW THz radiation and our THz comb mix on a receiver PCA, generating comb-CW beat signals (or beat notes). These signals are digitized by a 16-bit acquisition card (ATS9462, AlazarTech) at a sampling rate of 10 MS/s.

**Phase noise measurement.** For characterizing the phase noise of our near-infrared EO comb, we send it to a fast photodetector, whose output is measured by a commercial phase noise analyzer (ROHDE&SCHWARZ, 1 MHz-8 GHz). We perform comparison measurements (Fig. 1d) with and without the SSFS and spectral filtering stages, while keeping other measurement conditions unchanged. For phase noise measurements (Fig. 2d) of a THz comb-CW beat note, we first calculate its noise power spectral density (PSD) using the formula $PSD=P_{SSB}-10\lg(RBW)$ (dBm/Hz), where $P_{SSB}$ is the measured noise power at a given frequency, and RBW is the resolution bandwidth. We then obtain the phase noise using $PN = PSD\text{-}P_C$ (dBc/Hz), where $P_C = -42.1$ dBm is the carrier power of the beat note.

**Allan deviation analysis of THz comb lines.** For Allan deviation measurements, we use a digital frequency counter (FCA 3000, 1-s gate, Tektronix) to record a THz comb-CW beat signal over 30 minutes. The counter is referenced to the hydrogen maser. We then calculate fractional frequency instability by dividing the measured Allan deviation ($\sigma$) by the CW THz frequency ($f_{cw}$). This provides an upper limit on the frequency instability of our THz comb lines.

**THz time-domain spectroscopy.** In this experiment (Fig. 3a), twin near-infrared EO combs (EO combs 1 and 2) seeded by the same 1550-nm CW laser are used. One excites the transmitter PCA, generating THz comb radiation, while the other (incident power: 30 mW; pulse duration: 86 fs) electro-optically samples the THz waves on a receiver PCA (TERA 15-RX-FC, Menlo Systems). The two EO combs have slightly different repetition frequencies of 50 MHz and 49.999980 MHz, respectively. The photocurrent generated in the receiver PCA is converted into a voltage signal using a transimpedance amplifier (HCA-10M-100k, Femto), filtered by an electronic low-pass filter (cutoff frequency: 1.9 MHz), and then digitized by the acquisition card (sampling rate: 10 MS/s).